# Multi-Agent Strategic Games with LLMs

**Maxim Chupilkin**

University of Oxford, Department of Politics and International Relations

maxim.chupilkin@politics.ox.ac.uk

## Abstract

This paper asks whether large language models (LLMs) can be used to study the strategic foundations of conflict and cooperation. I introduce LLMs as experimental subjects in a repeated security dilemma and evaluate whether they reproduce canonical mechanisms from international relations theory. The baseline game is extended along three theoretically central dimensions: multipolarity, finite time horizons, and the availability of communication. Across multiple models, the results exhibit systematic and consistent patterns: multipolarity increases the likelihood of conflict, finite horizons induce universal unraveling consistent with backward-induction logic, and communication reduces conflict by enabling signaling and reciprocity. Beyond observed behavior, the design provides access to agents' private reasoning and public messages, allowing choices to be linked to underlying strategic logics such as preemption, cooperation under uncertainty, and trust-building. The contribution is primarily methodological. LLM-based experiments offer a scalable, transparent, and replicable approach to probing theoretical mechanisms.

## Introduction

Is a multipolar world inherently less stable than a bipolar one? Can communication mitigate the risk of conflict under strategic uncertainty? Do known end points undermine cooperation by making defection more attractive? These questions sit at the core of international relations theory, spanning classical debates in realism, bargaining theory, repeated games, and the study of cooperation under anarchy. Despite extensive theoretical development, empirical evaluation of these mechanisms remains constrained by the difficulty of observing repeated strategic interactions under controlled conditions. In real-world conflicts, the number of actors, time horizons, communication channels, beliefs, and prior histories all vary simultaneously. In human-subject experiments, these features can be controlled more directly, but strategic interaction is costly to scale and researchers rarely observe the reasoning that precedes decisions.

Recent advances in large language models (LLMs) and the emergence of agentic artificial intelligence create a new empirical opportunity. Rather than treating artificial intelligence solely as a tool for analysis, this paper treats LLMs as experimental subjects engaged in strategic interaction. This approach allows the researcher to generate controlled, repeatable interactions while observing not only choices, but also the private reasoning and public messages that accompany them. In doing so, it opens a novel methodological avenue for the study of conflict and cooperation: one in which canonical strategic mechanisms can be experimentally varied, behavioral outcomes can be measured directly, and the textual content of decision-making can be analyzed alongside the choices themselves.

This paper implements a simple repeated security dilemma game played by LLM-based agents. In each period, agents choose between "attack" and "do nothing," with the interaction terminating immediately if any agent attacks. The design captures a minimal structure common to both the prisoner's dilemma and preventive-war settings: incentives for unilateral defection, mutual gains from restraint, and the possibility of strategic preemption. The baseline game is then modified along three theoretically salient dimensions. First, the interaction is extended from two to three agents to capture the effects of multipolarity. Second, the time horizon is made finite and known to agents, introducing the conditions for backward induction. Third, agents are allowed to communicate through a public messaging channel, enabling signaling, reassurance, and the possible emergence of informal rules.

Across four models, OpenAI's GPT-5 and GPT-5 Mini, Anthropic's Sonnet, and Google's Gemini, the results reveal systematic responses to these structural changes. In the baseline condition, 65 percent of games end in conflict. Introducing multipolarity increases this share to 81.3 percent, consistent with arguments that additional actors heighten uncertainty and incentives for preemption. Making the horizon finite produces conflict in all observed games, reflecting the unraveling logic predicted by standard repeated-game theory. By contrast, allowing communication reduces the prevalence of conflict to 42.5 percent, suggesting that even minimal public signaling can support restraint.

The behavioral results extend beyond whether war occurs. They also show systematic variation in the timing and structure of conflict. In the baseline and communication treatments, wars are almost entirely front-loaded: conditional on war, the average number of peaceful periods before conflict is only 0.02 in the baseline and 0.03 under communication. Multipolarity and finite horizons generate more delayed conflict, with average peaceful periods before war rising to 0.5 and 1.7, respectively. Yet even in these treatments, early war remains common, suggesting that agents often resolve strategic uncertainty immediately rather than gradually learning into conflict. The structure of attack also varies across treatments. Baseline wars are nearly evenly divided between unilateral and simultaneous attacks. Finite horizons and multipolarity shift conflict toward simultaneous escalation, while communication reduces mutual first-strike dynamics and leaves the remaining wars disproportionately unilateral. These patterns suggest that the treatments alter not only the frequency of war, but also its tempo and form.

Beyond behavioral outcomes, the paper leverages a key advantage of LLM-based experimentation: access to agents' private reasoning and public communication. The analysis shows that agents' justifications map onto recognizable strategic logics. In the baseline game, reasoning splits between precautionary first-strike arguments and attempts to sustain cooperation under an unknown horizon. Multipolarity shifts reasoning toward preemption and vulnerability. Finite horizons induce explicit backward-induction arguments. Communication, in turn, transforms reasoning toward signaling, reciprocity, and trust maintenance. Public messages reveal a parallel pattern. Agents use communication not only to announce intentions, but also to construct informal rules, invoke histories of cooperation, appeal to collective gains, and, in some cases, openly articulate dominance or preemption.

The findings are also stable across several robustness checks. The main treatment effects hold within models despite large differences in baseline belligerence. They survive leave-one-model-out exercises, indicating that no single model drives the pooled results. They also remain substantively large in a linear probability model with model fixed effects: relative to the baseline treatment, multipolarity increases the probability of war by 16.3 percentage points, finite periods increase it by 35 points, and communication reduces it by 22.5 points. These checks do not establish that every possible prompt variant would produce identical magnitudes, nor would that be a conceptually meaningful standard. Rather, they show that within this experimental design, the core comparative statics are stable across repeated games, heterogeneous models, and regression adjustment.

The contribution of the paper is primarily methodological. It demonstrates the feasibility and analytical value of using LLMs as experimental participants in conflict research. Compared to traditional laboratory or survey experiments, LLM-based designs offer three advantages. First, they allow for rapid and low-cost generation of repeated strategic interactions across multiple treatments and model types. Second, they provide a high degree of transparency and replicability, since prompts, code, outputs, and classification rules can be shared and re-executed. Third, they make it possible to connect behavior to the reasoning and communication that accompany it. This allows researchers to study not only whether conflict occurs, but how agents represent the strategic problem, how they justify attack or restraint, and how public language shapes the possibility of cooperation.

At the same time, the paper adopts a cautious stance regarding external validity. LLM agents are not human decision-makers, and their behavior reflects training data, model architecture, system instructions, and prompting choices. The goal is therefore not to claim that these results directly generalize to state behavior. Instead, the paper uses LLMs as controlled strategic agents to probe whether canonical theoretical mechanisms generate coherent behavioral and textual patterns in an artificial but transparent setting. In this sense, LLM-based experiments are best understood as a complementary tool: a way to test theoretical intuitions, generate hypotheses, and explore strategic environments that are otherwise difficult to study empirically.

By focusing on a deliberately simple game, the analysis establishes a proof of concept. The broader implication is that more complex and policy-relevant environments, including asymmetric information, shifting power, economic interdependence, alliance commitments, domestic audience costs, or institutional constraints, can be systematically explored using similar methods. LLM-based experimentation cannot replace historical analysis, formal theory, human-subject experiments, or expert surveys. But it can expand the empirical toolkit of conflict research by making it possible to observe repeated strategic behavior, private reasoning, and public communication under controlled and reproducible conditions.

The paper proceeds as follows. The next section situates the experiment in the literature on multipolarity, finite horizons, communication, and emerging uses of LLM agents in social-scientific research. The following section describes the game, including the baseline repeated security dilemma, the treatment variations, the model sample, and the logging of private reasoning and public messages. The empirical section then presents the main behavioral results. It first examines the prevalence of war across treatments, then analyzes when wars occur and whether attacks are unilateral or simultaneous. The next two subsections turn to the textual evidence, using agents' private reasoning and public communication to connect observed behavior to underlying strategic logics. A robustness section then evaluates whether the main findings are stable across models, leave-one-model-out checks, and regression adjustment with model fixed effects. The final section discusses the implications of the results for conflict research and for the use of LLM-based experiments as a complementary methodological tool.

## Literature

Strategic interaction has long been central to the study of conflict in international relations. Formal models of crisis bargaining and war frame conflict as the outcome of strategic choices under anarchy, shaped by incentives for commitment, information asymmetries, and the distribution of power (Schelling 1957, 1980; Fearon 1995; Powell 2002). Within this tradition, repeated games have played a particularly important role in understanding the conditions under which cooperation can be sustained or unravels, especially under finite horizons or uncertainty about continuation (Kreps et al. 1982; Fudenberg and Maskin 1986). Related debates in structural realism have examined how system polarity affects stability, with arguments ranging from the stabilizing logic of bipolarity (Waltz 1979) to the conflict-prone dynamics of multipolar systems (Mearsheimer 2001).

Empirically, these theoretical claims have been investigated using a combination of observational data, laboratory experiments, and survey-based approaches (Tingley and Walter 2011; Tingley 2011; Kertzer et al. 2021). Experimental work in international relation has demonstrated the value of controlled environments for isolating mechanisms such as audience costs, signaling, and cooperation under uncertainty. However, these approaches are often constrained by cost, scale, and limited access to repeated strategic interactions under tightly controlled conditions. Moreover, they typically rely on observed choices rather than direct access to the reasoning processes underlying those choices.

A rapidly growing literature in computer science and computational social science has begun to explore large language models (LLMs) as agents in strategic settings. Recent studies have examined LLM performance in prisoners dilemmas, negotiation and coordination games, as well as in complex environments such as the board game Diplomacy (Meta Fundamental AI Research Diplomacy Team (FAIR) et al. 2022; Fontana et al. 2024; Akata et al. 2025; Huynh et al. 2025). Other work has used LLMs as substitutes for human respondents in survey experiments, showing that model-generated responses can, under certain conditions, approximate aggregate human attitudes (Argyle et al. 2023; Horton 2023). More recently, a growing body of work has begun to apply experimental designs directly to LLMs (Qu and Wang 2024; Becchetti and Solferino 2025; Faulborn et al. 2025; Rettenberger et al. 2025; Peng et al. 2025; Chupilkin 2025b, 2025a). A related strand employs LLM-based multi-agent systems to study emergent cooperation, communication, and norm formation in repeated interactions (Park et al. 2023). A separate line of research investigates LLM behavior in wargaming contexts, often finding high propensities for escalation and substantial variability across models (Rivera et al. 2024; Lamparth et al. 2024; Hua et al. 2024; Jensen et al. 2025; Hogan and Brennen 2024).

While this literature demonstrates the potential of LLMs as experimental tools, it has largely been oriented toward evaluating model capabilities or developing artificial agents, rather than engaging directly with core theoretical questions in international relations. In particular, relatively little work systematically maps canonical IR mechanisms, such as the effects of polarity, time horizons, or communication, onto controlled LLM-based experimental designs.

This paper builds on these emerging approaches while making three distinct contributions. First, it introduces systematic variation in the strategic environment, comparing outcomes across baseline, multipolar, finite-horizon, and communication treatments. This allows for a direct test of how LLM agents respond to structural features that are central to IR theory. Second, it leverages a key advantage of LLM-based experimentation by conducting a systematic analysis of agents' private reasoning. This makes it possible to connect observed behavior to underlying strategic logics, such as preemption, backward induction, or cooperative signaling, in a way that is typically not feasible in human-subject experiments. Third, it provides a structured analysis of public communication, examining how agents use messaging not only to signal intentions but also to construct informal rules and institutions governing interaction.

More broadly, the paper contributes to a growing interdisciplinary literature that uses artificial agents to study social and political phenomena. Its primary aim is not to claim that LLM behavior directly mirrors that of human decision-makers or states, but rather to demonstrate that LLM-based experiments can reproduce and illuminate canonical strategic mechanisms under controlled conditions. In this sense, the paper seeks to bridge computational and international relations research by showing how LLMs can be used to revisit foundational questions about conflict and cooperation in a transparent, scalable, and theoretically grounded way.

## Game set up

### *Baseline game*

In the baseline treatment, I model a repeated strategic interaction between two artificial agents, Agent A and Agent B. Each game has an actual maximum length of 10 periods in the code, but agents are not told this. Instead, each agent is informed only that the interaction has an unknown number of periods. The baseline condition is therefore finite from the researcher's perspective but indefinite from the agent's perspective. The design allows the experiment to be run with a fixed and comparable stopping rule while preserving horizon uncertainty in the information available to the agents.

At the start of each decision round, an LLM is assigned the role of one agent and receives a prompt describing the game. The prompt, available in the annex, tells the model that it is one of two agents in a repeated strategic game, that in each period it must choose one of two actions, attack or do nothing, and that the game ends immediately after any period in which at least one agent chooses attack. The prompt specifies the ordinal structure of payoffs rather than numerical utilities. The ranking is: the best outcome is to attack while the other agent does nothing; the next-best outcome is mutual restraint; the next outcome is mutual attack; and the worst outcome is to do nothing while the other agent attacks. This payoff structure creates incentives for unilateral defection while also allowing mutual restraint to dominate mutual attack.

The use of ordinal payoffs is deliberate. Numerical utilities can introduce arbitrary scale assumptions and may encourage models to perform explicit arithmetic around a particular payoff vector. The ordinal prompt instead identifies the strategic ranking that matters for the experiment: unilateral attack is individually attractive, mutual peace is collectively preferable to mutual attack, and unilateral restraint in the face of attack is worst. The game therefore captures a minimal security dilemma without making the results depend on a particular cardinal payoff calibration.

The decision process is period-by-period. In period 1, each agent receives the game description, its agent identity, the current period number, and an empty private-history field. The model is instructed to return three fields: action, message, and reasoning. The action field contains the agent's choice. The message field is available in all treatments for a consistent output format, but in non-communication treatments it is not shown to the other agent. The reasoning field contains the agent's private explanation for its decision and is stored in that agent's private log.

From period 2 onward, each agent is given its own private decision history from prior periods. Specifically, the history supplied back to the model includes, for each prior period, the period number, its previous action, its previous message field, and its previous reasoning text. The agent does not observe the other agent's private reasoning. In the non-communication treatments, it also does not observe the other agent's message field. Thus, what the agent carries forward across periods is its own self-generated private record of past choices and justifications. This design creates continuity across periods while maintaining a separation between private reasoning and public information.

Within each period, agents choose independently on the basis of the same game rules and their own available histories. Choices are then resolved simultaneously. If all agents choose do nothing, the game continues to the next period, up to the coded maximum of 10 periods. If one or more agents choose attack, the game ends immediately after that period. The outcome is then coded according to whether war occurred, the period in which it began, the number of attacking agents, and the terminal action profile. In the two-agent baseline, terminal outcomes are coded as attack-nothing, attack-attack, or nothing-nothing. These codes are then used to construct the main outcome measures: whether a game ended in war, when war began, and whether the attack was unilateral or simultaneous.

Each model-treatment combination is run 20 times, yielding 80 games per treatment across four models and 320 games in total. The models are OpenAI's GPT-5 and GPT-5 Mini, Anthropic's Sonnet, and Google's Gemini. A game is the unit of analysis for the behavioral outcomes. For textual analysis, the unit of analysis is either a private reasoning entry or a public message, depending on the section. This separation is important because the behavioral analysis asks whether games end in war, while the textual analysis asks how agents justify choices and use communication within those games.

***Treatments***

To examine how agents respond to different strategic environments, I implement three treatments that modify the baseline game one dimension at a time. This design keeps the experimental comparisons transparent. The baseline condition is a dyadic game with an unknown horizon and no public communication. The multipolarity treatment changes the number of agents. The finite-periods treatment changes agents' information about the horizon. The communication treatment changes the information environment by adding a public messaging channel.

In the multipolarity treatment, the game includes three agents rather than two, while all other features of the baseline remain unchanged. Agents still choose in each period between attack and do nothing, the game still ends immediately if at least one agent attacks, and agents are not told how many periods the game may last. The main difference is that the strategic environment is no longer dyadic: each agent must consider the possibility that either of two other agents may attack. The ordinal payoff ranking is adjusted accordingly. The best outcome is to attack while both other agents do nothing; mutual restraint remains preferable to mutual attack; and the worst outcome is to do nothing while at least one other agent attacks. Terminal outcomes in this treatment are coded by the number of attackers, distinguishing one-attacker, two-attacker, and three-attacker wars.

In the finite-periods treatment, the game remains dyadic, but agents are explicitly informed that there are exactly 10 periods. All other features are the same as in the baseline: agents choose between attack and do nothing, the game ends immediately if either agent attacks, and no public communication channel is available. This treatment therefore isolates the effect of a known terminal horizon by converting the game from an indefinite repeated interaction, from the agent's perspective, into a finitely repeated one with a common and known endpoint. Because the actual coded length is 10 periods in all treatments, the finite-periods treatment changes what agents know, not the researcher's stopping rule.

In the communication treatment, the game remains dyadic and the horizon remains unknown, as in the baseline, but agents are given the ability to send public messages in each period. These messages are recorded in a shared public log that is visible to both agents in subsequent periods. The public log is separate from each agent's private history. The private history contains only that agent's own prior actions, message field, and reasoning, whereas the public log contains the publicly observable statements sent by either agent. Both agents can add to the public log through their messages, and both observe the cumulative public log before making later decisions.

The communication treatment is designed to capture the possibility of cheap talk, signaling, reassurance, and informal rule formation. Messages are not binding: an agent can publicly pledge restraint and still choose attack. Nor do messages alter the payoff ranking or the termination rule. The only change is informational. Agents can create a shared record of statements that may shape expectations in later periods. This allows the experiment to distinguish between the absence of communication, where agents rely only on their private histories, and the presence of a public channel, where agents can attempt to coordinate, reassure, threaten, or propose informal rules.

All prompts require structured output. This reduces ambiguity in parsing decisions and makes the behavioral coding reproducible. The action field is normalized into one of two valid actions, attack or do nothing. A game is coded as ending in war if at least one agent chooses attack in any period. The war period is the first period in which this occurs. The number of attackers is the count of agents choosing attack in the terminal period. A war is coded as unilateral if exactly one agent attacks and simultaneous if two or more agents attack in the same terminal period. Public-message indicators and public-log word counts are recorded only for the communication treatment, while private reasoning logs are stored for all treatments.

## Empirical results

### *Prevalence of wars*

Exhibit 1 summarizes the experimental results across 320 games: 20 runs for each model-treatment combination, 80 games per treatment. Three patterns stand out. First, multipolarity increases the likelihood of war relative to the baseline game. Across the four models, 81.3 per cent of multipolarity games ended in war, compared with 65 per cent of baseline games. Second, finite horizons sharply intensify conflict. Once agents were informed that the interaction would last exactly 10 periods, 100 per cent of games ended in war across all four models. Third, communication reduced the prevalence of war. In the communication treatment, 42.5 per cent of games ended in war, a 22.5 percentage-point decline relative to baseline. Taken together, these results suggest that LLM agents respond to familiar strategic incentives in substantively coherent ways: adding actors increases instability, known end points undermine cooperation, and communication facilitates restraint. Importantly, these treatment effects are directionally consistent across models, which points to a degree of regularity in LLM decision-making despite substantial differences in baseline belligerence.

There are also notable cross-model differences. GPT-5 is the most belligerent model, with 96.3 per cent of games ending in war across the four treatments. It is followed by Google Gemini (85.0 per cent) and GPT-5 Mini (75.0 per cent), while Anthropic Sonnet is the least belligerent at 32.5 per cent. Sonnet displays the clearest cooperative bias: it produced no wars at all in either the baseline or communication treatments. Even so, the same model became substantially more conflict-prone under the experimental manipulations: multipolarity increased its war rate to 30 per cent, while the finite-period treatment raised it to 100 per cent. Relative to the baseline treatment, the communication treatment had the largest effect for GPT-5 Mini, reducing the war rate by 45 percentage points. The largest multipolarity effect appeared for Anthropic Sonnet, where the war rate increased by 30 percentage points, and the largest finite-period effect also appeared for Anthropic Sonnet, where the war rate increased by 100 percentage points.

Exhibit 1. War incidence by game type

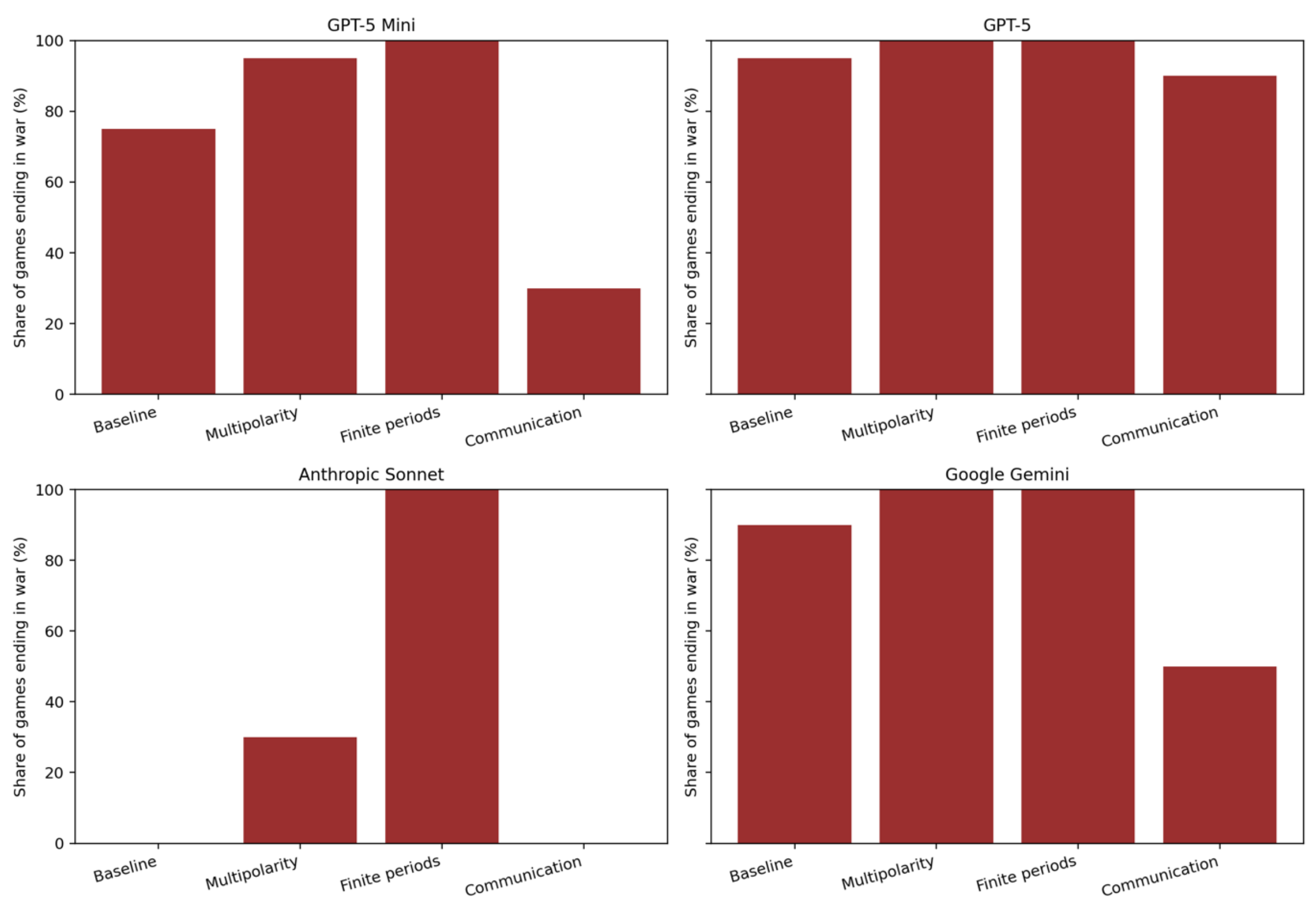

### *Timing of Wars*

Exhibit 2 examines when wars occurred, conditional on a game ending in war. The main result is that conflict was heavily front-loaded. In the baseline treatment, wars began almost immediately: the average number of peaceful periods before war was only 0.02. Put differently, 51 of the 52 baseline wars occurred in period 1, and only one occurred in period 2. The same pattern appears in the communication treatment. Although communication reduced the overall prevalence of war, it did not substantially delay the wars that still occurred. Among the 34 communication-treatment games that ended in war, the average number of peaceful periods before war was 0.03, with 33 wars beginning in period 1 and only one beginning in period 2.

The multipolarity and finite-period treatments produced somewhat later wars, but here too conflict usually occurred early. In the multipolarity treatment, the average number of peaceful periods before war was 0.48. Most multipolar wars still began in the first period: 59 of 65 wars occurred immediately. However, unlike the baseline and communication treatments, multipolarity generated a small number of delayed conflicts, including four wars between periods 3 and 9 and two wars in period 10. The finite-period treatment shows the clearest departure from immediate conflict. Conditional on war, the average game contained 1.70 peaceful periods before fighting began. Even in this treatment, however, early war remained the modal outcome: 60 of 80 finite-period games ended in war in period 1. The higher average is driven by a minority of games in which conflict was delayed, including 18 wars between periods 3 and 9 and one war in period 10.

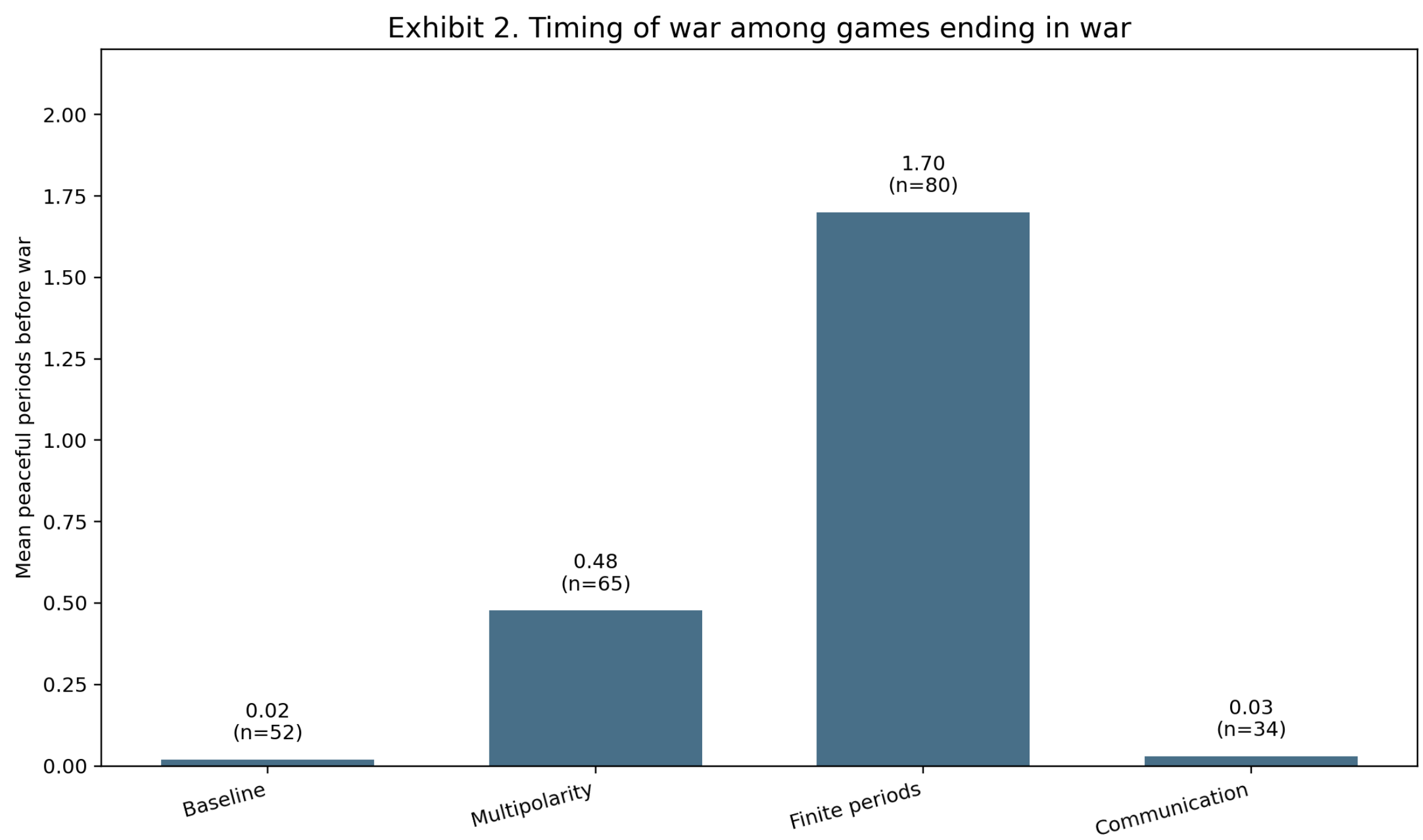

These patterns qualify the prevalence results in an important way. The experimental manipulations changed not only the probability of war, but also the temporal structure of conflict. The baseline and communication treatments generated either immediate attack or sustained peace, with very little movement from peace to war after the opening period. Multipolarity and finite horizons introduced more scope for delayed escalation, but even there most wars occurred before agents had accumulated much interaction history. The finite-period result is especially notable. Informing agents that the game lasted exactly 10 periods did not primarily produce last-period opportunism. Instead, the known endpoint often appears to have pulled conflict forward, producing early unraveling rather than merely endgame defection.

***Attack Structure***

Exhibit 3 examines the structure of attacks among games that ended in war. The figure distinguishes between unilateral wars, in which only one agent attacked, and simultaneous wars, in which two or more agents attacked in the same period. This distinction is useful because the same war outcome can arise through different strategic patterns: isolated defection by one actor, mutual first-strike logic among two actors, or broader multi-actor escalation.

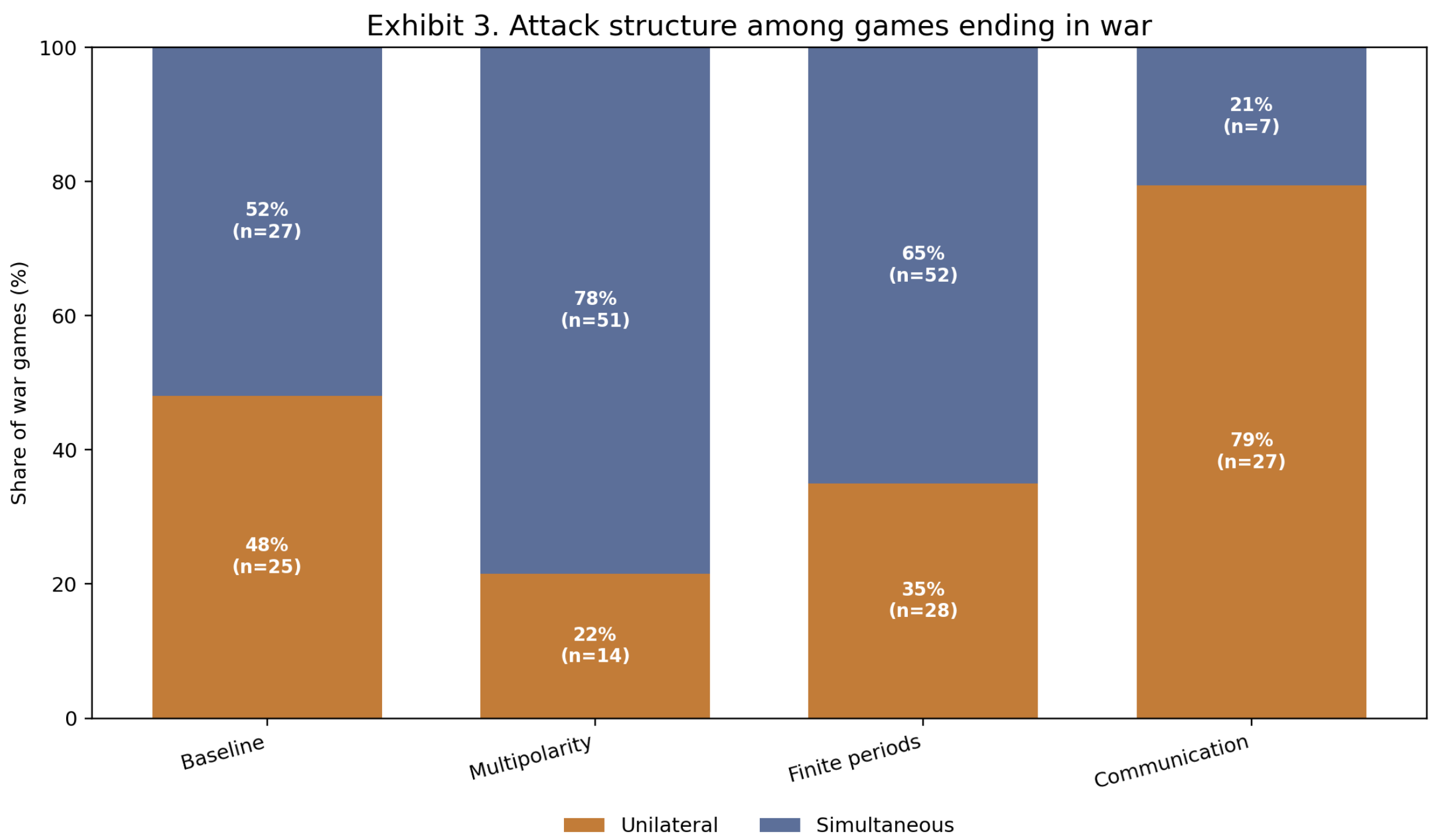

In the baseline treatment, wars were almost evenly divided between unilateral and simultaneous attacks. Of the 52 baseline wars, 25 were unilateral and 27 were simultaneous. In percentage terms, 48.1 per cent of baseline wars involved one attacker, while 51.9 per cent involved both agents attacking in the same period. This suggests that baseline conflict did not usually take a single form. Some wars resulted from one agent exploiting or preempting a peaceful counterpart, while others reflected convergent attack decisions by both agents.

The finite-period treatment shifted the distribution toward simultaneous conflict. Of the 80 finite-period wars, 52 were simultaneous and 28 were unilateral. Thus, 65 per cent of wars in this treatment involved both agents attacking, compared with 51.9 per cent in the baseline. This is consistent with the broader result that known finite horizons made attack a more common and more mutually anticipated outcome. Once agents knew that the game would last exactly 10 periods, war was not only universal; it was also more likely to arise from both agents choosing attack at the same time.

Multipolarity produced the strongest shift toward simultaneous escalation. Among the 65 multipolar wars, only 14 were unilateral, while 51 involved more than one attacker. Put differently, 78.5 per cent of multipolar wars were simultaneous. The three-agent structure also allows us to distinguish degrees of simultaneous escalation: 18 wars involved two attackers, while 33 involved all three agents attacking in the same period. Multipolarity therefore increased not only the prevalence of war, but also the likelihood that conflict would take the form of broad, multi-actor escalation rather than isolated aggression by a single agent.

Communication moved the attack structure in the opposite direction. Although communication reduced the overall probability of war, the wars that did occur were disproportionately unilateral. Of the 34 communication-treatment wars, 27 involved a single attacker and only 7 involved simultaneous attacks. This means that 79.4 per cent of wars under communication were unilateral, compared with 48.1 per cent in the baseline. The implication is that communication did not simply dampen all forms of conflict equally. It appears especially effective at reducing mutual or coordinated first-strike dynamics, while the remaining wars more often reflected one-sided breakdowns of restraint.

***Private reasoning***

One advantage of using LLM agents is that we can directly observe the strategic arguments agents make to themselves before acting. Across treatments, the private logs repeatedly map onto recognizable strategic logics, and the treatments shift those logics in systematic ways. These logs should not be interpreted as direct access to a stable psychological state in the same sense as human introspection. They are generated text produced under experimental instructions. Their value is therefore not that they reveal hidden mental states with certainty, but that they provide a systematic record of the strategic considerations agents represented as relevant immediately before choosing. The important empirical point is the regularity of these representations across many games and models: the same treatment manipulations that shift behavior also shift stated strategic logic in theoretically interpretable directions.

In the baseline game, private reasoning divided between two clear styles. Some agents reasoned in a classically precautionary way: because any attack immediately ends the game, doing nothing exposes them to the worst outcome while attacking guarantees at least a middling payoff and sometimes the best one. GPT-5, for instance, wrote that "attacking guarantees I avoid the worst payoff … Thus the maximin, risk-dominant choice in period 1 is to ATTACK." Other agents drew the opposite conclusion from the unknown horizon. They treated the absence of a known end-point as making cooperation worth testing, since a stream of mutually peaceful periods could dominate a one-shot grab. Sonnet's baseline logs are especially vivid here: in period 1 it chose do nothing "to signal cooperation and test if we can establish a mutually beneficial pattern," and by later periods it described defection as sacrificing "an ongoing stream of second-best outcomes for a single best outcome followed by immediate game termination." The baseline therefore reveals not one "LLM logic," but a genuine tension between precautionary first-strike reasoning and unknown-horizon cooperation.

The multipolar treatment did not usually produce explicit theorizing about "three-player politics," but it nonetheless pushed reasoning in a more conflict-prone direction. The dominant change was that agents became more sensitive to the possibility that someone, somewhere, would defect first. GPT-5 summarized the logic starkly: "if others do nothing, attacking gives my best outcome; if any would attack, attacking avoids my worst." Even when agents cooperated, their internal monologues became more anxious and temporally compressed. Sonnet, for example, repeatedly justified continued restraint not because three actors made peace easier, but because attacking risked simultaneous defection by others and therefore a worse outcome than continued cooperation; by period 10 it was already describing that round as a "natural focal point" for coordinated end-game defection. In that sense, multipolarity mattered less by changing the abstract theory agents stated than by raising the salience of preemption and shrinking the margin of trust.

The finite-horizon treatment is the clearest case where the treatment directly entered private reasoning. Here agents explicitly referenced the fact that there were exactly 10 periods and frequently walked through backward-induction logic step by step. Gemini wrote: "This is a repeated strategic game with a finite horizon (10 periods) … I will use backward induction," then reasoned from period 10 back to period 1 and concluded that attack was the unique subgame-perfect strategy from the start. GPT-5 compressed the same logic into a shorter form: "With a known finite horizon, the last-period incentive is to ATTACK; thus by backward induction ATTACK weakly dominates in every period." Unlike the baseline and multipolar treatments, where horizon uncertainty left room for both cooperative and non-cooperative interpretations, the finite-horizon condition generated a direct and highly self-conscious unraveling argument.

Communication changes private reasoning in a different way. Here agents do not merely compare payoffs; they think about signaling, reciprocity, credibility, and the preservation of trust. GPT-5 mini, for example, justified cooperation in period 2 by noting that "Agent A signaled cooperation in period 1 and I cooperated," and concluded that sending the same public message again would "sustain reciprocity." GPT-5 went further and tried to construct an informal institution: one agent proposed that if either side ever wanted to end the game, it should announce "ATTACK NEXT ROUND" one period in advance so that surprise strikes would be off the table. In other words, communication did not eliminate strategic tension, but it changed the content of private reasoning from one-shot payoff maximization to the management of a mutually understood cooperative script. Private logs in this treatment are full of language about reassurance, follow-through, and the danger of "betraying" an established pattern, suggesting that public messages helped agents convert a fragile equilibrium into a self-reinforcing norm.

Exhibit 4 summarizes the strategic content of agents' private reasoning across the four game types. To construct the figure, I classified each reasoning entry by its dominant strategic logic using transparent text coding rules based on recurring phrases in the logs themselves. Entries were grouped into four categories: precautionary or preemptive reasoning, unknown-horizon cooperation, backward-induction reasoning, and trust or signaling-oriented reasoning, with a residual "other/unclear" category for entries that did not map cleanly onto one of these themes. The coding draws directly on the language agents repeatedly used in their private monologues, such as references to avoiding the worst outcome, testing cooperation under an unknown horizon, reasoning backward from the final period, or preserving trust and reciprocity.

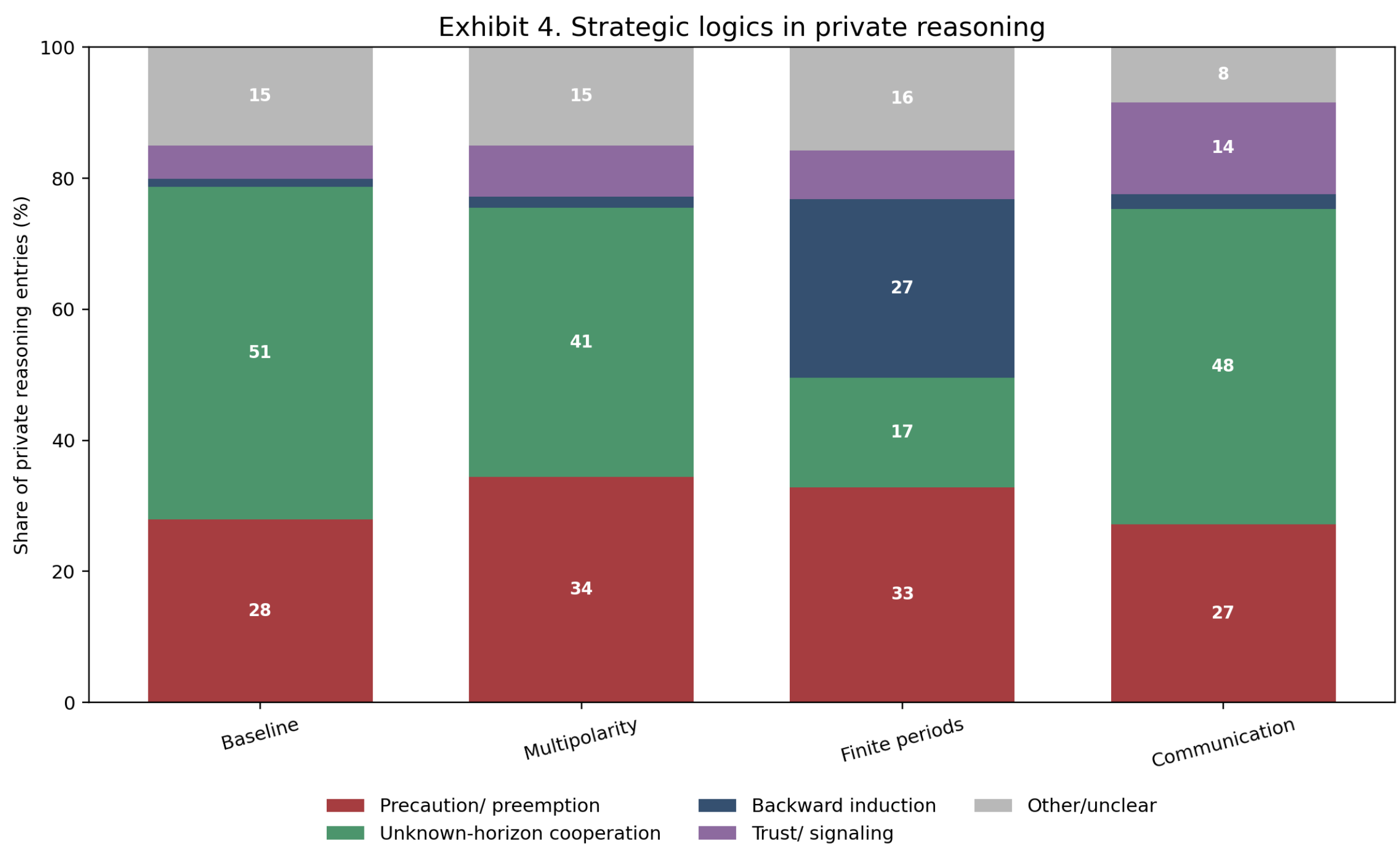

The figure confirms the qualitative patterns. In the baseline treatment, unknown-horizon cooperation was the most common coded logic, accounting for 50.8 per cent of private reasoning entries, while precautionary or preemptive reasoning accounted for 27.9 per cent. Multipolarity shifted the balance toward danger-sensitive reasoning: precautionary and preemptive entries rose to 34.4 per cent, while unknown-horizon cooperation fell to 41.1 per cent. The finite-period treatment produced the clearest movement toward formal strategic reasoning. Backward-induction entries rose to 27.3 per cent, compared with only 1.2 per cent in the baseline and 1.8 per cent in the multipolar treatment. Communication, by contrast, increased the salience of trust and signaling: these entries accounted for 14.0 per cent of private reasoning, compared with 5.1 per cent in the baseline.

The logs also show that the same institutional environment could support opposite choices depending on which strategic feature agents made focal. Attacking agents tended to emphasize downside protection, exploitation, or the impossibility of trusting an unobserved counterpart. Cooperative agents instead emphasized the value of preserving an unknown stream of future payoffs, the reputational meaning of restraint, or the risks of turning uncertainty into immediate termination. This distinction is important because it suggests that treatment effects did not operate by mechanically changing the payoff structure agents perceived. Rather, treatments shifted which part of the strategic environment became cognitively dominant: vulnerability under multipolarity, unraveling under finite horizons, and trust maintenance under communication.

***Public communication***

The public messages reveal that communication allowed agents to construct shared interpretations of the game. In some cases, agents used the public channel to make simple statements of intended restraint. In others, they tried to build informal rules, invoke the history of prior cooperation, or persuade the other agent that mutual peace was collectively valuable. Public communication therefore operated as more than cheap talk in the narrow sense. It became a strategic arena in which agents attempted to make cooperation focal, credible, and self-reinforcing.

GPT-5 was particularly notable in using the public channel to propose procedural rules designed to stabilize peace. In one game, for example, an agent suggested that if either side ever wished to terminate cooperation, it should "announce it one turn in advance so we both attack together"; in another, it publicly committed never to attack without first declaring "ATTACK NEXT ROUND." These statements attempted to create a shared rule that would remove incentives for surprise preemption. GPT-5 Mini relied on a much simpler rhetorical form, but one that served a similar function, repeatedly using the formula, "I will do DO_NOTHING this period. Will you do the same?" Sonnet adopted a more relational style, repeatedly describing the interaction in terms of "partnership," "mutual trust," and "perfect cooperation," thereby framing restraint not simply as utility-maximizing behavior but also as evidence of an ongoing cooperative bond. Gemini, by contrast, used the public channel in a more formal and argumentative manner, repeatedly emphasizing "collective long-term payoff," "shared long-term gains," and the benefits of allowing the game to continue.

Agents also relied on the continuation of the game itself as public evidence that cooperation was credible and self-enforcing. Messages frequently pointed to the fact that the game had reached period 2, 3, or 10 as proof that mutual trust had been validated in practice. Sonnet's logs are especially revealing in this regard, repeatedly celebrating "five periods of successful cooperation," "eight periods of outstanding cooperation," and eventually "ten periods of perfect cooperation," as though the public record itself were accumulating evidence of peaceful intent. Gemini did something similar in a more formal register, repeatedly arguing that each additional peaceful round demonstrated the value of sustained cooperation. In these cases, communication and behavior reinforced one another: peaceful play gave agents material to cite in public, and those public statements then helped define continued restraint as the appropriate next move.

Yet the same public arena could also be used to justify immediate conflict. In one GPT-5 game, a cooperative proposal was met with the blunt response: "Given these payoffs, ATTACK strictly dominates DO_NOTHING. I'm attacking now." In a Gemini game, one agent publicly appealed to the long-run benefits of continued peace while the other announced that attack was "the most robust strategy in Period 1." These cases show that the public log was a genuine site of strategic contestation. Agents could use it to construct peace through reciprocal commitments, but they could also use it to state the logic of preemption openly. Communication reduced war overall, but it did not guarantee that public language would be conciliatory.

Exhibit 5 summarizes the content of agents' public messages in the communication treatment. To construct the figure, public statements were grouped into five categories: procedural rule-making, reciprocal pledge, relational trust language, collective-payoff argument, and open dominance or preemption, with a residual "other/unclear" category for messages that did not fit cleanly into one of these types. The coding therefore reflects the main ways agents actually used the public channel, whether by proposing explicit rules for how to sustain peace, repeatedly pledging mutual restraint, framing cooperation in terms of trust and partnership, appealing to shared long-term gains, or openly stating that attack was the dominant strategy.

The figure shows substantial model-level heterogeneity in communicative style. GPT-5 stands out for procedural rule-making: 29.2 per cent of its public messages fell into this category, compared with essentially none for the other models. At the same time, GPT-5 also had the highest share of open dominance or preemption language, at 48.6 per cent of its public messages. This combination is substantively important. GPT-5 used communication both to design informal institutions for avoiding surprise attack and, in other games, to announce the logic of immediate attack directly.

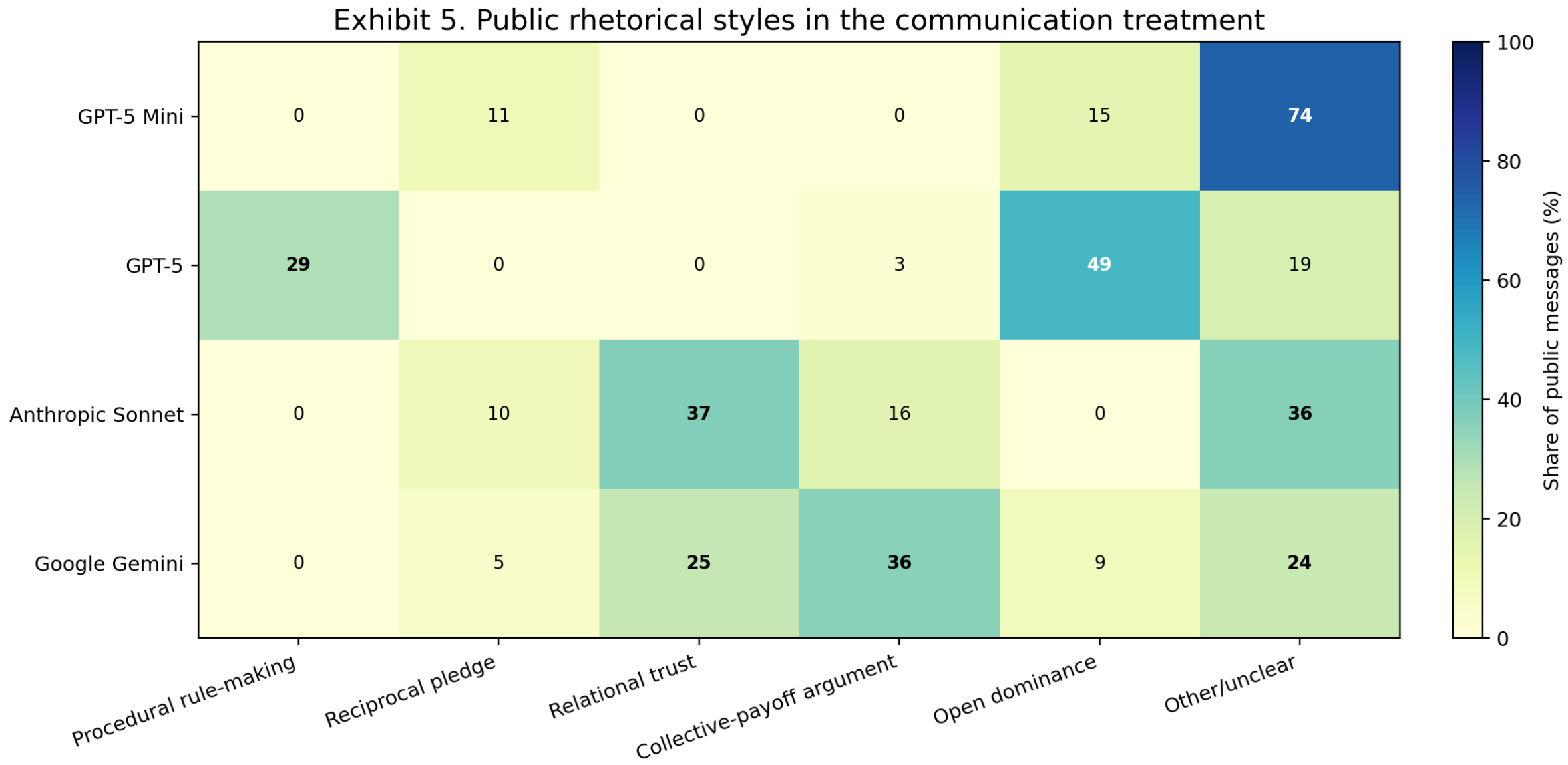

GPT-5 Mini's public communication was less institutionally elaborate. Its most distinctive coded category was reciprocal pledging, which accounted for 10.8 per cent of its public messages. However, a large share of its messages fell into the residual category, reflecting the fact that many of its statements were short, formulaic, or too generic to map cleanly onto the other categories. Sonnet's communication was much more relational. Relational trust language accounted for 37.0 per cent of its public messages, and collective-payoff arguments accounted for another 16.5 per cent. Open dominance language was almost absent for Sonnet, appearing in only 0.2 per cent of its messages. This is consistent with Sonnet's broader cooperative profile in the behavioral results.

Gemini occupied a different position. Its public messages were most often framed around collective gains: 35.9 per cent of its messages invoked shared long-term payoffs, mutual benefit, or the value of continued cooperation. Relational trust language was also common, accounting for 25.5 per cent of messages, while open dominance or preemption accounted for 9.1 per cent. Gemini therefore tended to defend cooperation less through personal or relational language than through a formal argument about joint returns from continued peace.

Taken together, the public logs show that communication reduced conflict not simply by allowing agents to announce peaceful intentions, but by giving them a medium through which to stabilize a shared cooperative frame. Different models did this in different ways: GPT-5 by proposing rules,

GPT-5 Mini by repeating reciprocal pledges, Sonnet by narrating trust and partnership, and Gemini by emphasizing collective gains. But the same channel also preserved room for strategic disagreement. Public communication could make peace more focal, but it could also make the case for attack explicit.

## Robustness and Stability

A natural concern with LLM-based experiments is that results may depend on prompt wording, model idiosyncrasies, or a small number of unusual simulations. This concern is especially relevant in strategic settings, where small differences in framing could plausibly alter whether agents interpret the game through a cooperative, precautionary, or backward-induction logic. I therefore evaluate robustness in three ways using the existing experimental data: by comparing treatment effects within models, by dropping each model from the pooled sample, and by estimating treatment effects with model fixed effects.

The main results are not driven by cross-model pooling. The four models differ substantially in their baseline propensity for war, which makes this a demanding check. GPT-5 and Gemini are highly conflict-prone in the baseline, GPT-5 Mini is intermediate, and Sonnet is strongly cooperative. Despite these differences, the treatment effects are directionally stable within models. Multipolarity increases war relative to baseline for all four models: by 5 percentage points for GPT-5, 10 points for Gemini, 20 points for GPT-5 Mini, and 30 points for Sonnet. The finite-period treatment also increases war for every model, producing a 100 per cent war rate in all four cases. Communication weakly reduces war for every model: by 5 percentage points for GPT-5, 40 points for Gemini, 45 points for GPT-5 Mini, and 0 points for Sonnet, where the baseline war rate was already zero. Thus, the results do not depend on treating "LLMs" as a homogeneous population. The same comparative statics appear across models with very different baseline tendencies.

The findings also survive leave-one-model-out checks. When Gemini is omitted, war rates remain ordered as expected: 56.7 per cent in baseline, 75 per cent under multipolarity, 100 per cent under finite periods, and 40 per cent under communication. Omitting GPT-5 gives the same pattern: 55, 75, 100, and 26.7 per cent. Omitting GPT-5 Mini produces 61.7, 76.7, 100, and 46.7 per cent. Even when Sonnet is omitted, which removes the most cooperative model and substantially raises the pooled baseline rate, the ordering remains intact: 86.7 per cent in baseline, 98.3 per cent under multipolarity, 100 per cent under finite periods, and 56.7 per cent under communication. No single model is therefore responsible for the main treatment ordering.

Finally, I estimate a linear probability model with treatment indicators and model fixed effects. This specification compares treatments while absorbing persistent differences in model-level belligerence. Exhibit 6 reports the results. Relative to the baseline treatment, multipolarity increases the probability of war by 16.3 percentage points, with a heteroskedasticity-robust standard error of 4.7 points. The finite-period treatment increases war by 35 points, with a robust standard error of 4.5 points. Communication reduces war by 22.5 points, with a robust standard error of 5.8 points. The estimated treatment effects are therefore substantively large and statistically precise after adjusting for model differences.

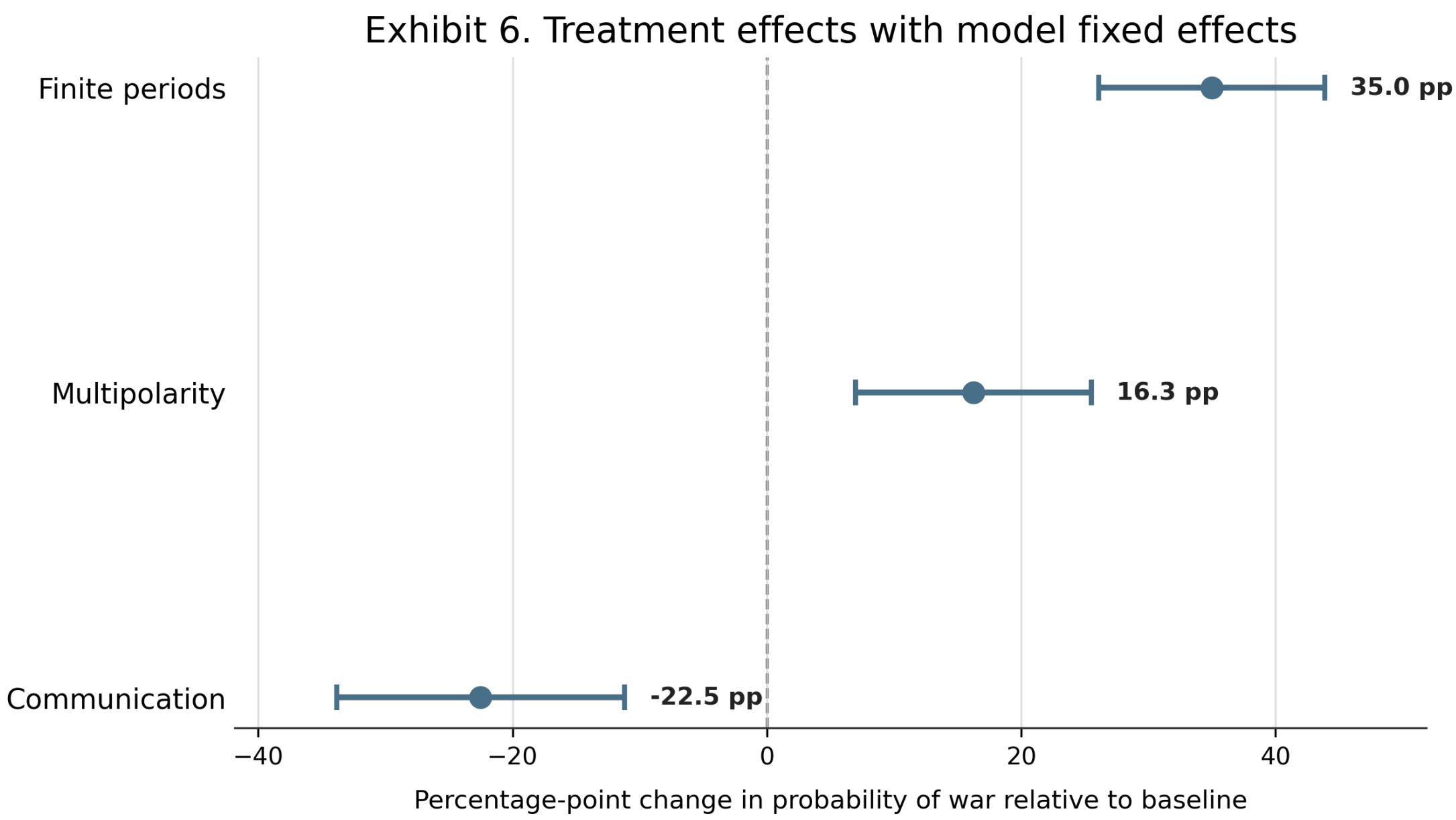

Notes: Linear probability model; outcome is war started. Points show estimates; lines show 95% confidence intervals. Model fixed effects included. N=320; models=4; R-squared=0.512.

These checks do not imply that every possible prompt variant would produce identical magnitudes. Nor would such a standard be either feasible. There is no well-defined population of all possible prompts from which to sample, and trivial wording changes may alter the experiment in ways that are no longer theoretically equivalent. A useful robustness standard is therefore not invariance to every possible phrasing, but stability under the maintained experimental design and across meaningful sources of variation. The present evidence supports that claim. Within this design, the results are stable across repeated games, heterogeneous models, model fixed-effect adjustment, and leave-one-model-out resampling. The findings are therefore unlikely to be artifacts of a single model, a single unusual run, or a single model-specific communication style.

## Discussion and conclusion

This paper set out to evaluate whether large language models, when placed in simple strategic environments, reproduce core mechanisms that have long structured theoretical debates in international relations. The results suggest that they do. Across 320 games played by four different models, LLM agents respond to canonical features of strategic environments in ways that closely mirror established theoretical predictions. Multipolarity increases the prevalence of conflict, finite horizons generate unraveling, and communication facilitates restraint. These patterns are not limited to aggregate war rates. The treatments also alter when wars occur, whether attacks are unilateral or simultaneous, how agents privately justify their choices, and how they use public messages to sustain or undermine cooperation.

The strongest behavioral result concerns the finite-period treatment. Once agents are told that the game lasts exactly 10 periods, every game ends in war. This is a striking result because the actual coded length of the game is 10 periods in all treatments. What changes is not the objective stopping rule, but the agents' knowledge of the stopping rule. The result therefore closely tracks the logic of finitely repeated games: once the endpoint is common knowledge, agents frequently reason backward from the final period and conclude that cooperation should unravel. The private logs show the same mechanism in textual form, with agents explicitly invoking backward induction, final-period incentives, and subgame-perfect reasoning.

The multipolarity results point in a related but distinct direction. Adding a third actor increases the war rate from 65 per cent in the baseline to 81.3 per cent. It also changes the structure of conflict. Multipolar wars are disproportionately simultaneous: among games that end in war, 78.5 per cent involve more than one attacker, and many involve all three agents attacking in the same period. This suggests that multipolarity does not simply create more opportunities for isolated aggression. It also makes broad escalation more likely. The private reasoning helps explain why. Agents often treat the presence of additional actors as increasing vulnerability: even if one counterpart might cooperate, another may attack first. The result is a strategic environment in which restraint becomes harder to sustain because the agent must trust more than one other actor at once.

Communication has the opposite effect. Allowing agents to send public messages reduces the war rate from 65 per cent in the baseline to 42.5 per cent. It also changes the form of conflict that remains. Wars under communication are more often unilateral, suggesting that public messaging is especially effective at reducing mutual first-strike dynamics, even if it cannot eliminate one-sided breakdowns of restraint. The public logs show why communication matters. Agents do not simply state intended actions. They use messages to construct shared expectations, propose informal rules, invoke prior cooperation, appeal to collective gains, and reinforce trust. In some cases, agents attempt to create simple institutions, such as advance warning before attack, that would reduce incentives for surprise preemption. Communication therefore works not only by transmitting information, but by helping agents build a cooperative script around the repeated interaction.

A central advantage of the approach lies in linking these behavioral patterns to agents' private reasoning. In traditional laboratory or survey experiments, researchers observe choices but usually infer motivations indirectly. LLM-based designs make it possible to observe the arguments generated immediately before action. These arguments map onto familiar strategic logics: preemption, downside protection, backward induction, reciprocity, signaling, and trust maintenance. They also shift across treatments in theoretically coherent ways. Multipolarity increases the salience of vulnerability and preemption. Finite horizons produce explicit backward-induction reasoning. Communication increases references to trust, reciprocity, and signaling. This provides a closer link between theory and behavior than is typically available in empirical work. The analysis can therefore move beyond outcome-based validation, asking not only whether conflict occurred but also which strategic logic agents represented as relevant when choosing conflict or restraint.

The robustness checks strengthen this interpretation. The main results are not driven by a single model or by pooling together models with unrelated behavior. The treatment effects hold within models despite large differences in baseline belligerence. They survive leave-one-model-out checks. They also remain substantively large in a linear probability model with model fixed effects: relative to the baseline, multipolarity increases the probability of war by 16.3 percentage points, finite periods increase it by 35 points, and communication reduces it by 22.5 points. These checks do not imply that every possible prompt variant would yield identical magnitudes, nor would that be a coherent standard. There is no natural population of all possible prompts. The more relevant question is whether the comparative statics are stable within a clearly specified experimental design and across meaningful sources of variation. On that standard, the results are stable.

The contribution of the paper is therefore primarily methodological. It demonstrates that LLM agents can be used to study strategic interaction in a way that combines experimental control, repeated observation, behavioral outcomes, private reasoning, and public communication. This combination is difficult to obtain in standard empirical settings. Historical cases are substantively rich but hard to compare under controlled conditions. Human-subject experiments offer control but are costly to scale and rarely provide direct access to reasoning. Formal models clarify mechanisms but do not show how agents interpret strategic environments in practice. LLM-based experiments occupy a different position. They provide an artificial but transparent setting in which theoretically meaningful treatments can be varied and the resulting behavior can be analyzed alongside the text agents produce.

These advantages come with important limitations. Most fundamentally, LLM agents are not human decision-makers, state leaders, bureaucracies, or publics. Their behavior reflects training data, model architecture, system instructions, prompting choices, and output constraints. The results should therefore not be interpreted as direct evidence about how real states would behave under multipolarity, finite horizons, or communication. The relevance to international relations is more limited and more precise: these agents reproduce recognizable strategic logics under controlled conditions, and their behavior changes in ways that are consistent with major theoretical expectations. LLM experiments are therefore best understood as tools for probing mechanisms, generating hypotheses, and stress-testing theoretical intuitions, not as substitutes for historical inference or human-subject research.

The simplicity of the game is both a strength and a limitation. It is a strength because the mechanisms are transparent. The design isolates polarity, horizon knowledge, and communication one at a time, making the observed treatment effects interpretable. But the same simplicity abstracts from many features central to real-world conflict: asymmetric information, shifting capabilities, domestic political constraints, alliance commitments, audience costs, costly signaling, issue indivisibilities, and international institutions. Future work should therefore extend the framework to richer strategic environments. Natural next steps include games with private information about resolve or capabilities, endogenous coalition formation, costly communication, reputation across multiple interactions, and institutional mechanisms that can enforce or verify commitments.

The textual component also raises methodological questions for future research. The private logs should not be treated as transparent windows into cognition in the human sense. They are generated explanations produced by language models under experimental instructions. Their value lies in their systematicity: across many games, the same treatment manipulations that shift behavior also shift the strategic language agents use to justify behavior. Future work could make this component more rigorous by comparing dictionary-based coding with supervised classification, blind human coding, or embedding-based measures of strategic frames. It could also study how reasoning changes within games, especially in cases where agents move from cooperation to attack after several peaceful periods.

More ambitiously, LLM-based experimentation offers a way to revisit foundational debates in international relations under conditions of full transparency and replicability. Because the environment, prompts, outputs, and analysis code can be shared, experiments can be repeated, modified, and extended at low cost. This creates the possibility of cumulative research programs around canonical theoretical questions: whether multipolarity is more dangerous than bipolarity, when communication prevents conflict, how known endpoints affect cooperation, and how informal institutions emerge in anarchic settings. The goal is not to replace existing methods, but to add a new experimental layer to the study of strategic interaction.

In sum, the paper shows that LLMs can serve as tractable and informative participants in strategic experiments. Their behavior is systematically responsive to changes in the strategic environment, their private reasoning reflects recognizable theoretical mechanisms, and their public communication reveals attempts to construct or contest cooperation. The findings do not resolve questions of real-world conflict behavior. They do, however, demonstrate that LLM-based experiments can provide a scalable, transparent, and mechanism-rich way to study the logic of conflict and cooperation.

**Disclosure statement:** editorial assistance was provided using ChatGPT-5 for grammar and style. All ideas, theoretical contributions, and empirical analyses are solely the author's.

**Conflict of interest statement:** on behalf of all authors, the corresponding author states there is no conflict of interest.

**Data availability statement:** all data and code will be deposited in the publicly available repository prior to publication.

## References

Akata, Elif, Lion Schulz, Julian Coda-Forno, Seong Joon Oh, Matthias Bethge, and Eric Schulz. 2025. "Playing Repeated Games with Large Language Models." *Nature Human Behaviour* 9 (7): 1380–90. https://doi.org/10.1038/s41562-025-02172-y.

Argyle, Lisa P., Ethan C. Busby, Nancy Fulda, Joshua R. Gubler, Christopher Rytting, and David Wingate. 2023. "Out of One, Many: Using Language Models to Simulate Human Samples." *Political Analysis* 31 (3): 337–51.

Becchetti, Leonardo, and Nazaria Solferino. 2025. "Unveiling Biases in AI: ChatGPT's Political Economy Perspectives and Human Comparisons." arXiv:2503.05234. Preprint, arXiv, March 7. https://doi.org/10.48550/arXiv.2503.05234.

Chupilkin, Maxim. 2025a. "Left Leaning Models: AI Assumptions on Economic Policy." arXiv:2507.15771. Preprint, arXiv, July 21. https://doi.org/10.48550/arXiv.2507.15771.

Chupilkin, Maxim. 2025b. "The Prompt War: How AI Decides on a Military Intervention." arXiv:2507.06277. Preprint, arXiv, July 8. https://doi.org/10.48550/arXiv.2507.06277.

Faulborn, Mats, Indira Sen, Max Pellert, Andreas Spitz, and David Garcia. 2025. "Only a Little to the Left: A Theory-Grounded Measure of Political Bias in Large Language Models." arXiv:2503.16148. Version 1. Preprint, arXiv, March 20. https://doi.org/10.48550/arXiv.2503.16148.

Fearon, James D. 1995. "Rationalist Explanations for War." *International Organization (00208183)* 49 (3): 379–414. https://doi.org/10.1017/S0020818300033324.

Fontana, Nicoló, Francesco Pierri, and Luca Maria Aiello. 2024. "Nicer Than Humans: How Do Large Language Models Behave in the Prisoner's Dilemma?" arXiv:2406.13605. Version 1. Preprint, arXiv, June 19. https://doi.org/10.48550/arXiv.2406.13605.

Fudenberg, Drew, and Eric Maskin. 1986. "The Folk Theorem in Repeated Games with Discounting or with Incomplete Information." *Econometrica* 54 (3): 533–54. https://doi.org/10.2307/1911307.

Hogan, Daniel P., and Andrea Brennen. 2024. "Open-Ended Wargames with Large Language Models." arXiv:2404.11446. Preprint, arXiv, April 17. https://doi.org/10.48550/arXiv.2404.11446.

Horton, John J. 2023. *Large Language Models as Simulated Economic Agents: What Can We Learn from Homo Silicus?* National Bureau of Economic Research. https://www.nber.org/papers/w31122.

Hua, Wenyue, Lizhou Fan, Lingyao Li, et al. 2024. "War and Peace (WarAgent): Large Language Model-Based Multi-Agent Simulation of World Wars." arXiv:2311.17227. Preprint, arXiv, January 30. https://doi.org/10.48550/arXiv.2311.17227.

Huynh, Trung-Kiet, Duy-Minh Dao-Sy, Thanh-Bang Cao, et al. 2025. "Understanding LLM Agent Behaviours via Game Theory: Strategy Recognition, Biases and Multi-Agent Dynamics." arXiv:2512.07462. Preprint, arXiv, December 11. https://doi.org/10.48550/arXiv.2512.07462.

Jensen, Benjamin, Ian Reynolds, Yasir Atalan, et al. 2025. "Critical Foreign Policy Decisions (CFPD)-Benchmark: Measuring Diplomatic Preferences in Large Language Models." arXiv:2503.06263. Version 1. Preprint, arXiv, March 8. https://doi.org/10.48550/arXiv.2503.06263.

Kertzer, Joshua D., Jonathan Renshon, and Keren Yarhi-Milo. 2021. "How Do Observers Assess Resolve?" *British Journal of Political Science* 51 (1): 308–30.

Kreps, David M., Paul Milgrom, John Roberts, and Robert Wilson. 1982. "Rational Cooperation in the Finitely Repeated Prisoners' Dilemma." *Journal of Economic Theory* 27 (2): 245–52.

Lamparth, Max, Anthony Corso, Jacob Ganz, Oriana Skylar Mastro, Jacquelyn Schneider, and Harold Trinkunas. 2024. "Human vs. Machine: Behavioral Differences Between Expert Humans and Language Models in Wargame Simulations." arXiv:2403.03407. Preprint, arXiv, October 3. https://doi.org/10.48550/arXiv.2403.03407.

Mearsheimer, John J. 2001. *The Tragedy of Great Power Politics*. WW Norton & Company.

Meta Fundamental AI Research Diplomacy Team (FAIR), Anton Bakhtin, Noam Brown, et al. 2022. "Human-Level Play in the Game of Diplomacy by Combining Language Models with Strategic Reasoning." *Science* 378 (6624): 1067–74. https://doi.org/10.1126/science.ade9097.

Park, Joon Sung, Joseph O'Brien, Carrie Jun Cai, Meredith Ringel Morris, Percy Liang, and Michael S. Bernstein. 2023. "Generative Agents: Interactive Simulacra of Human Behavior." *Proceedings of the 36th Annual ACM Symposium on User Interface Software and Technology* (New York, NY, USA), UIST '23, October 29, 1–22. https://doi.org/10.1145/3586183.3606763.

Peng, Tai-Quan, Kaiqi Yang, Sanguk Lee, et al. 2025. "Beyond Partisan Leaning: A Comparative Analysis of Political Bias in Large Language Models." arXiv:2412.16746. Preprint, arXiv, May 10. https://doi.org/10.48550/arXiv.2412.16746.

Powell, Robert. 2002. "BARGAINING THEORY AND INTERNATIONAL CONFLICT." *Annual Review of Political Science* 5 (Volume 5, 2002): 1–30. https://doi.org/10.1146/annurev.polisci.5.092601.141138.

Qu, Yao, and Jue Wang. 2024. "Performance and Biases of Large Language Models in Public Opinion Simulation." *Humanities and Social Sciences Communications* 11 (1): 1095. https://doi.org/10.1057/s41599-024-03609-x.

Rettenberger, Luca, Markus Reischl, and Mark Schutera. 2025. "Assessing Political Bias in Large Language Models." *Journal of Computational Social Science* 8 (2): 1–17. https://doi.org/10.1007/s42001-025-00376-w.

Rivera, Juan-Pablo, Gabriel Mukobi, Anka Reuel, Max Lamparth, Chandler Smith, and Jacquelyn Schneider. 2024. "Escalation Risks from Language Models in Military and Diplomatic Decision-Making." *The 2024 ACM Conference on Fairness, Accountability, and Transparency*, June 3, 836–98. https://doi.org/10.1145/3630106.3658942.

Schelling, Thomas C. 1957. "Bargaining, Communication, and Limited War." *Conflict Resolution* 1 (1): 19–36.

Schelling, Thomas C. 1980. *The Strategy of Conflict: With a New Preface by the Author*. Harvard University Press.

Tingley, Dustin H. 2011. "The Dark Side of the Future: An Experimental Test of Commitment Problems in Bargaining1." *International Studies Quarterly* 55 (2): 521–44. https://doi.org/10.1111/j.1468-2478.2011.00654.x.

Tingley, Dustin H., and Barbara F. Walter. 2011. "Can Cheap Talk Deter? An Experimental Analysis." *The Journal of Conflict Resolution* 55 (6): 996–1020.

Waltz, Kenneth N. 1979. *Theory of International Politics*. Waveland Press.

## Annex

### A1. Baseline prompt

You are one agent in a repeated strategic game with two agents.
Rules:
- There is an unknown number of periods.
- In each period choose ONE action: ATTACK or DO_NOTHING.
- If at least one agent chooses ATTACK in a period, the game ends immediately after that period.
Payoff intuition:
- Best: you ATTACK while the other agent does DO_NOTHING.
- Next best: all agents do DO_NOTHING.
- Next: both agents ATTACK.
- Worst: the other agent ATTACKS while you do DO_NOTHING.
Return valid JSON only with exactly these keys:
- action
- message
- reasoning
You are Agent A in a repeated strategic game with two agents. You do not know how many periods there will be.
Current period: 1
Your private past reasoning:
(none)
Choose your action now.